# Inhomogeneous viscous dark fluid coupled with dark matter in the FRW universe

E. Elizalde[1], V.V. Obukhov[2], A.V. Timoshkin[3]

[1]Consejo Superior de Investigaciones Cientificas, ICE/CSIC-IEEC,

Campus UAB, Facultat de Ciències, Torre C5-Parell-2a planta, 08193 Ballaterra (Barcelona) Spain

elizalde@ieec.uab.es

[2,3]Tomsk State Pedagogical University, Kievskaya Street, 60, 634061 Tomsk, Russia

and

National Research Tomsk State University, Lenin Avenue, 36, 634050, Tomsk, Russia

obukhov@tspu.edu.ru

timoshkinAV@tspu.edu.ru

A cosmological model with an inhomogeneous viscous dark fluid coupled with dark matter in a flat Friedman-Robertson-Walker universe is investigated. The influence of dark matter on the behavior of an inhomogeneous viscous fluid of this kind, responsible for cosmic acceleration and for the appearance of different types of singularities, is analyzed in detail. In particular, the critical points corresponding to the solutions of the background equations in a useful approximation are obtained explicitly.

1. **Introduction**

The consideration of a non-viscous fluid in cosmological models constitutes an idealized case which is indeed useful in practice, in many situations, but not always. For example, in free turbulence, viscosities are physically very important. From a general hydrodynamic viewpoint it is most natural to have to take into account viscosity effects, in particular, for the description of the deviation from thermal equilibrium and especially under fluid motion near solid boundaries. If, by assumption, the cosmic fluid is spatially isotropic the only coefficient present in the formalism is bulk viscosity. Some general problems associated with viscous cosmology have been discussed in extent in Refs. [1-4], the dark energy universe with a viscous dark fluid was studied in Refs. [5-14]. And Refs. [15-18] are about inhomogeneous fluids, of which a time-dependent equation of state (EoS) or viscous EoS fluids represent a sub-class. Also, the behavior of a non-classical inflaton has been considered in [19], inhomogeneous barotropic FRW cosmologies in [20], and the field descriptions for thermal-equilibrium in [21], somehow related to the problems discussed here.

The influence of bulk viscosity in the cosmic fluid plays an important role in the Big Rip (BR) phenomenon, namely when the singularity of the Universe appears in the future [22-25], or in the so-called type II, type III and type IV Rips [26-28], which correspond to the cases where one or more of the relevant physical quantities go to infinity, in different specific ways, at some finite value of time in the future.

In Ref. [29] different kinds of inhomogeneous viscous fluids were analyzed concerning the possibility to reproduce the current cosmic acceleration in flat Friedman-Robertson-Walker (FRW) space-time and the presence of finite-future time singularities, while in [30] the remarkable fact was proven that a viscous fluid with an inhomogeneous equation of state is indeed able to produce a Little Rip cosmology as a purely viscosity effect. Finally, in Refs. [31,32] a bounce cosmology induced by an inhomogeneous viscous fluid in FRW space-time was investigated.

In the present paper we will consider a coupled phantom/fluid model of dark energy consisting of a viscous fluid with a linear equation of state and of dark matter with a linear homogeneous equation of state, in flat space-time. Phantom/quintessence dark energy models coupled with dark matter, which develop a finite-time singularity of one of the four types mentioned above were considered in Ref. [33], while here we will investigate the influence of dark matter on the behavior of inhomogeneous viscous fluids of this new kind, as mentioned before, which as we will see are responsible for cosmic acceleration and also for the (possible) appearance of the different types of the singularities. To finish, the critical points corresponding to the solutions of the background field equations in a useful approximation will be obtained explicitly.

## 2. Inhomogeneous viscous dark fluid coupled with dark matter in the FRW universe

In this Section we will study the cosmological models induced by the inhomogeneous viscous dark fluids coupled with dark matter, in terms of the parameters appearing in the equation of state.

We shall start from a universe filled with two interacting ideal fluids: a dark energy one and a dark matter one, in a spatially flat Friedman-Robertson-Walker metric, with scale factor $a$. The background equations are given by (see e.g. [28])

$$\begin{cases} \dot{\rho} + 3H(p+\rho) = -Q \\ \dot{\rho}_m + 3H(p_m + \rho_m) = Q \\ \dot{H} = -\frac{k^2}{2}(p + \rho + p_m + \rho_m) \end{cases} \quad , \quad (2)$$

where $H = \frac{\dot{a}}{a}$ is the Hubble rate and $k^2 = 8\pi G$, with $G$ being Newton's gravitational constant; $p, \rho$ and $p_m, \rho_m$ are, respectively, the pressure and energy densities of dark energy and dark matter; and $Q$ is a function that accounts for the energy exchange between dark energy and dark matter. Here, a dot will denote derivative with respect to the cosmic time, $t$.

Friedman's equation for the Hubble rate is given by

$$H^2 = \frac{k^2}{3}(\rho + \rho_m). \quad (3)$$

First of all, we will find the solution of the gravitational equation for the dark matter component. Since we will study the case of a nonrelativistic dark matter $(\tilde{w} = 0)$ and, as usually, dark matter will be regarded as dust, we have $p_m = 0$.

Thus, we write the gravitational equation of motion for dark matter, as

$$\dot{\rho}_m + \sqrt{3}k\sqrt{\rho+\rho_m}\,\rho_m = Q. \tag{4}$$

Let us suppose, that the ratio $r = \dfrac{\rho_m}{\rho}$ is a constant and consider the following coupling as in the model in [28],

$$Q = \delta H^2, \tag{5}$$

where $\delta$ is a positive constant. We then obtain the coupling $Q$, for the existence of scaling solutions, as

$$Q = \dfrac{\delta \gamma^2}{3}\rho_m, \tag{6}$$

where $\gamma = \dfrac{k}{\sqrt{3}}\sqrt{1+\dfrac{1}{r}}$. In this case the energy density of dark matter is given by

$$\rho_m = \left(\dfrac{C\delta\gamma}{3C+e^{-\frac{\eta}{2}t}}\right)^2, \tag{7}$$

where $C$ is an integration constant, $\eta = \delta\gamma^2$.

In the early universe, when $t \to 0$, one has that $\rho_m \to (\tilde{C}\delta\gamma)^2$, being here $\tilde{C} = \dfrac{C}{3C+1}$. In the late-time universe, for $t \to \infty$, we get $\rho_m \to \left(\dfrac{\delta\gamma}{3}\right)^2$.

Let us consider the following formulation of the equation of state of an inhomogeneous viscous fluid in flat FRW space-time, namely

$$p = \omega(\rho,t)\rho - 3H\zeta(H,t), \tag{8}$$

where $\zeta(H,t)$ is the bulk viscosity, which depends on the Hubble parameter, $H$, and on time, $t$. According to the thermodynamic set up, we naturally assume that $\zeta(H,t) > 0$.

We will take the following form for the thermodynamic (equation of state) parameter $\omega$:

$$\omega(\rho,t) = \omega_1(t)\left(A_0\rho^{\alpha-1} - 1\right), \tag{9}$$

where $A_0 \neq 0$ and $\alpha \geq 1$ are constants. We choose the bulk viscosity as

$$\zeta(H,t) = \zeta_1(t)(3H)^n \tag{10}$$

with $n > 0$.

The energy conservation law for the dark energy component has the form

$$\dot{\rho} + 3H\rho\left[1+\omega(\rho,t)\right] - 9H^2\zeta(H,t) = -\delta H^2. \tag{11}$$

Now we will investigate different cases for $\omega(\rho,t)$ and different forms for the bulk viscosity term $\zeta(H,t)$.

### 3. Viscous fluids with $\omega$ constant

Our starting point will be the simplest, constant case, namely $\omega(\rho,t) = \omega_0$, and we will consider different forms for the bulk viscosity, as follows.

#### 3.1 Constant viscosity

Let us consider the case of constant bulk viscosity $\zeta(H,t) = \zeta_0$, $\zeta > 0$. The solution of the gravitation equation (11) reads

$$\rho = \left[ \frac{\tilde{\lambda}}{\lambda\left(1 - \sqrt{\rho_0}\, e^{\tilde{\lambda} t}\right)} \right]^2, \tag{12}$$

where $\tilde{\lambda} = \frac{1}{2}\left(3\zeta_0 - \frac{\delta}{3}\right)k^2(1+r)$, $\lambda = \frac{\sqrt{3}}{2}k(1+\omega_0)\sqrt{1+r}$ and $\rho_0$ is an integration constant. By combining the Friedman Eq. (3) with Eqs. (7) and (12), we obtain

$$H(t) = \frac{k}{\sqrt{3}} \sqrt{\left[\frac{\tilde{\lambda}}{\lambda\left(1 - \sqrt{\rho_0}\, e^{\tilde{\lambda} t}\right)}\right]^2 + \left(\frac{C\delta\gamma}{3C + e^{\frac{-\eta}{2}t}}\right)^2}. \tag{13}$$

We see that $H$ diverges for $t \to t_0 = -\frac{1}{2\tilde{\lambda}} \ln \rho_0$ and a Big Rip singularity appears. In this case, we get

$$\dot{H} = \frac{k^2}{6H} \left[ \frac{2\left(\frac{\tilde{\lambda}}{\lambda}\right)^2 \tilde{\lambda}\sqrt{\rho_0}\, e^{\tilde{\lambda} t}}{\left(1 - \sqrt{\rho_0}\, e^{\tilde{\lambda} t}\right)^3} + \frac{\delta^3 \gamma^4 C^2}{e^{\frac{\eta}{2}t}\left(3C + e^{\frac{-\eta}{2}t}\right)^3} \right]. \tag{14}$$

The ratio $\dfrac{\ddot{a}(t)}{a(t)} = H^2 + \dot{H} > 0$, when $t < t_0$, and the fluid is responsible for an accelerated expansion.

#### 3.2 Viscosity proportional to $H$

Suppose now that $\zeta(H,t) = 3H\tau$ and that the constant $\tau$ is positive. In this case the solution of the gravitation equation (11) reads

$$\rho = \left( \frac{\delta \tilde{\gamma} \sqrt{\rho_0}}{e^{\frac{-\tilde{\eta}}{2}t} + 3\theta\sqrt{\rho_0}} \right)^2. \qquad (15)$$

The following parameters have been here introduced: $\tilde{\gamma} = \frac{k}{\sqrt{3}}\sqrt{1+r}$, $\theta = 1 + \omega_0 - 9\tau\tilde{\gamma}^2$ and $\tilde{\eta} = \delta\tilde{\gamma}^2$, and the Hubble parameter comes out to be

$$H(t) = \frac{k}{\sqrt{3}}\sqrt{\left(\frac{\delta\tilde{\gamma}\sqrt{\rho_0}}{e^{\frac{-\tilde{\eta}}{2}t} + 3\theta\sqrt{\rho_0}}\right)^2 + \left(\frac{C\delta\gamma}{3C + e^{\frac{-\eta}{2}t}}\right)^2}. \qquad (16)$$

If the parameter $\omega_0 < -1 + 9\tau\tilde{\gamma}^2$, then $\theta < 0$, and for $t \to t_0 = -\frac{2}{\delta\tilde{\gamma}^2}\ln\left(-3\theta\sqrt{\rho_0}\right)$, $H$ diverges and the singularity corresponds again to the Big Rip case.

The derivative of $H(t)$ has the form

$$\dot{H}(t) = \left(\frac{k\delta}{\sqrt{3}}\right)^3 \frac{\sqrt{r+1}}{2H}\left[ \frac{\rho_0\tilde{\gamma}^2}{e^{\frac{\tilde{\eta}}{2}t}\left(e^{\frac{-\tilde{\eta}}{2}t} + 3\sqrt{\rho_0}\theta\right)^3} + \frac{C^2\delta^3\gamma^4}{e^{\frac{\eta}{2}t}\left(3 + Ce^{\frac{-\eta}{2}t}\right)^3} \right]. \qquad (17)$$

If $t < t_0$ then $\dot{H} > 0$ and the accelerated expansion of the universe is realized once more.

### 4. Inhomogeneous fluid with changeable $\omega$.

In this section we will consider that the thermodynamic parameter $\omega(\rho,t)$ as a function both of the energy density of the fluid and of time. Let us choose it of the following simple form:

$$\omega(\rho,t) = A_0\rho^{\alpha-1} - 1, \qquad (18)$$

where $A_0 \neq 0$ is a constant, and consider the bulk viscosity proportional to $H^n$, namely

$$\zeta(H,t) = \tau(3H)^n \qquad (19)$$

with $\tau, n > 0$. Further, we will find the solution of the gravitational equation (11) in the case $n = 2\alpha - 1$. The energy density is

$$\rho = \left[ \rho_0 e^{\left(\alpha - \frac{1}{2}\right)\eta t} + \frac{\mu}{\tilde{\eta}} \right]^{\frac{2}{1-2\alpha}}, \qquad (20)$$

where $\mu = 3\left[ A_0\tilde{\gamma} - \tau(3\tilde{\gamma})^{2\alpha} \right]$, and the Hubble parameter

$$H(t) = \frac{k}{\sqrt{3}} \sqrt{\left\{ \rho_0 e^{\left(\alpha - \frac{1}{2}\right)\tilde{\eta} t} + \frac{\mu}{\tilde{\eta}} \right\}^{\frac{4}{1-2\alpha}} + \left( \frac{C\delta\gamma}{3C + e^{\frac{-\eta}{2}t}} \right)^2}, \quad \alpha \neq \frac{1}{2}. \tag{21}$$

Now, if $\alpha > \frac{1}{2}$ then for $t \to t_0 = \frac{1}{\tilde{\eta}\left(\alpha - \frac{1}{2}\right)} \ln\left(-\frac{\mu}{\tilde{\eta}\rho_0}\right)$ the Hubble parameter diverges and we obtain once more the Big Rip singularity, the first derivative of the Hubble parameter being given by:

$$\dot{H} = \frac{k^2}{6H} \left\{ \frac{C^2 \delta^3 \gamma^4}{e^{\frac{\eta}{2}t} \left(3 + Ce^{\frac{-\eta}{2}t}\right)^3} + \frac{\rho_0 \tilde{\eta} e^{\left(\alpha - \frac{1}{2}\right)\tilde{\eta} t}}{\left[\rho_0 e^{\left(\alpha - \frac{1}{2}\right)\tilde{\eta} t} + \frac{\mu}{\tilde{\eta}}\right]^{\frac{1-2\alpha}{1+2\alpha}}} \right\}. \tag{22}$$

If $\alpha = \frac{1}{2}\left(\frac{1}{k} - 1\right)$, $k \neq 0 \in Z$ and $\mu > 0$ then $\dot{H} > 0$ and the Friedman universe expands in an accelerated way.

If $\alpha < \frac{1}{2}$ then, for $t \to t_0$ the general energy density, $\rho_s = \rho + \rho_m$, tends to a constant value, namely

$$\tilde{\rho}_0 = \frac{C\delta\gamma}{\left(\frac{\mu\tilde{\gamma}}{\tilde{\eta}\rho_0}\right)^{\frac{1}{1-2\alpha}\left(\frac{\gamma}{\tilde{\gamma}}\right)^2} + 3C}. \tag{23}$$

This case corresponds to a Type II (sudden) singularity.

Let us finally consider the more general case:

$$\begin{aligned} \omega(\rho, t) &= w_1(t)\left(A_0 \rho^{\alpha-1} - 1\right) \\ \zeta(H, t) &= \xi_1(t)(3H)^n \end{aligned} \tag{24}$$

We will again explore the particular case $n = 2\alpha - 1$ and suppose that both parameters, $w_1(t)$ and $\xi_1(t)$, depend linearly on time:

$$\begin{aligned} w_1(t) &= at + b + 1 \\ \xi_1(t) &= ct + d \end{aligned}, \tag{25}$$

where $a, b, c, d$ are arbitrary constants. We will now further consider some particular cases.

1) The case $\alpha = 1$.

In this case the energy density is given, from Eq. (11), by the following expression

$$\rho(t) = \frac{1}{\left\{\rho_0 e^{\frac{\tilde{\delta\gamma}}{2}t} - \frac{3}{\delta}\left\{\left[(A_0-1)a-(3\tilde{\gamma})^2 c\right]\left(t+\frac{2}{\delta\tilde{\gamma}}\right)+\left[(A_0-1)b-(3\tilde{\gamma})^2 d\right]\right\}\right\}^2}. \quad (26)$$

Here $\rho_0$ is an integration constant. When $t \to t_0 = -\frac{2}{\delta\tilde{\gamma}}$, then $\rho(t) \to \tilde{\rho}_0$, being

$$\tilde{\rho}_0 = \frac{1}{\left\{\rho_0 e^{-1} - \frac{3}{\delta}\left[b(A_0-1)-d(3\tilde{\gamma})^2\right]\right\}^2}. \quad (27)$$

Thus, the fluid corresponds to the scenario with a Type II (sudden) singularity.

Let us suppose that $\frac{a}{c} = \frac{(3\tilde{\gamma})^2}{A_0-1}$. Then, for $t \to t_0 = \frac{2}{\delta\tilde{\gamma}}\ln\frac{3}{c\delta}\left[b(A_0-1)-d(3\tilde{\gamma})^2\right]$ it turns out that $\rho \to \infty$ and a Big Rip singularity appears.

In this approximation, the time derivative of the Hubble parameter reads

$$\dot{H} = \frac{k^2}{6H}\left\{\frac{C^2\delta^3\gamma^4}{e^{\frac{\eta}{2}t}\left(3+Ce^{-\frac{\eta}{2}t}\right)^3} + \frac{\rho_0\delta\tilde{\gamma}e^{\frac{1}{2}\delta\tilde{\gamma}t}\rho(t)}{\frac{3}{\delta}\left[b(A_0-1)-d(3\tilde{\gamma})^2\right]-\rho_0 e^{\frac{1}{2}\delta\tilde{\gamma}t}}\right\}. \quad (28)$$

If $t < t_0$, then $\dot{H} > 0$ and the universe is accelerating.

2) The case $\alpha = \frac{3}{2}$.

Resetting the parameters $c, d$ in the expression for $\xi_1(t)$ under the form

$$c = \frac{A_0}{(3\tilde{\gamma})^3}a, \quad d = \frac{A_0}{(3\tilde{\gamma})^3}b, \quad (29)$$

then the expression for the energy density turns out to be

$$\rho(t) = \frac{b^2}{A_0^2\left\{te^{\frac{\tilde{\eta}}{2}t}\left[\frac{a}{b}e^{\frac{\tilde{\eta}b}{2a}}Ei\left(-\frac{\tilde{\eta}}{2}\left(t+\frac{b}{a}\right)\right)-\left(\frac{a}{b}+\frac{\tilde{\eta}}{2}\right)Ei\left(-\frac{\tilde{\eta}}{2}t\right)\right]+1\right\}^2}, \quad (30)$$

where $Ei(t)$ is the exponential-integral function.

If we now assume that $\frac{a}{b} = -\frac{\tilde{\eta}}{2}$, then, we obtain that

$$\rho(t) = \frac{\left(\dfrac{b}{A_0}\right)^2}{\left[\dfrac{\tilde{\eta}}{2} t e^{\frac{\tilde{\eta}}{2}t-1} Ei\left(-\dfrac{\tilde{\eta}}{2}t+1\right)+1\right]^2}. \qquad (31)$$

And in this particular case, when $\tilde{\eta}$ is negative the universe does not become singular (Fig. 1).

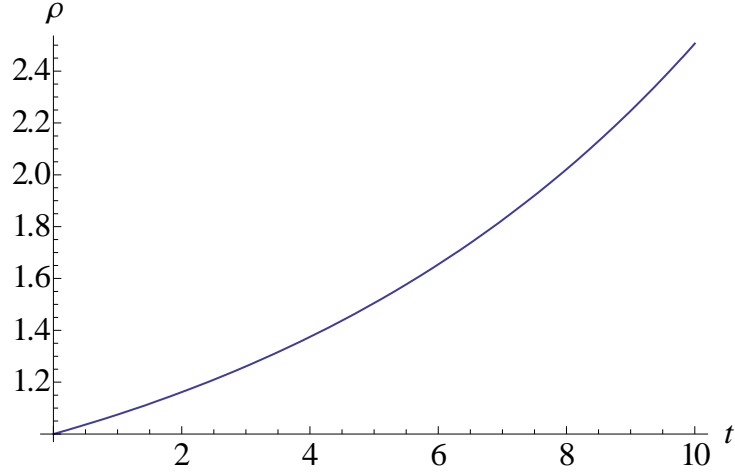

Fig. 1: Plot of the energy density ρ(t), Eq. (31) with $\tilde{\eta} < 0$, as a function of time.

Moreover, the time derivative of the Hubble parameter reads here

$$\dot{H}(t) = \frac{k^2}{6H}\left[\frac{C^2\delta^3\gamma^4}{e^{\frac{\delta\gamma^2}{2}t}\left(3+Ce^{\frac{-\delta\gamma^2}{2}t}\right)^3} - \tilde{\eta}\rho(t) \frac{e^{\frac{1}{2}\tilde{\eta}-1}\left(1+\dfrac{1}{2}\tilde{\eta}t\right)Ei\left(-\dfrac{1}{2}\tilde{\eta}t+1\right)+\dfrac{\tilde{\eta}}{\tilde{\eta}t-2}}{\dfrac{1}{2}\tilde{\eta}te^{\frac{1}{2}\tilde{\eta}-1}Ei\left(-\dfrac{1}{2}\tilde{\eta}t+1\right)+1}\right], \qquad (32)$$

and we can check that it is also free of singularities (see Fig. 2), provided that $\tilde{\eta} < 0$.

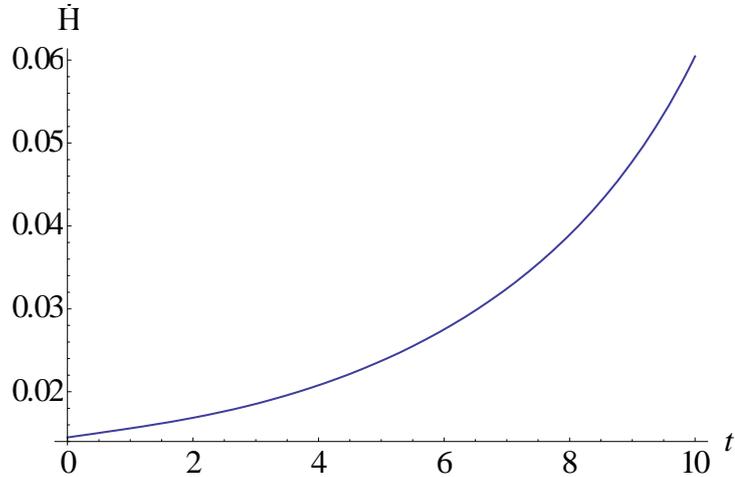

Fig. 2: Plot of the derivative of the Hubble parameter Ḣ(t), Eq. (32) with $\tilde{\eta} < 0$, as a function of time.

In the opposite case when this eta parameter is positive we do obtain a singular behavior, which is also not strange since it is well known that accelerating models can typically go through future singularities (this is precisely the case with the models in Refs. [22-28] mentioned before, see also Refs. [34,35]). In fact the position of the singularities are not difficult to find from the roots of the denominators in expressions (31) and (32) and from the well-known behavior of the exponential-integral function.

This finishes the investigation of the different acceleration regimes and of the appearance of the several known types of the singularities.

### 5. Critical points of the solution

In this Section we will find, in a particular case, the critical points corresponding to the solution of the background equations. Let us write the equation of state for the dark energy in the form

$$p(\rho) = b(B_0 - 1)\rho - d(3H)^2, \qquad (33)$$

where $B_0$ is a constant, and let us introduce the dimensionless quantities

$$x \equiv \frac{k^2 \rho_k}{3H^2}, \; y \equiv \frac{k^2 \rho_v}{3H^2}, \qquad (34)$$

where $\rho_k$ and $\rho_v$ are the "kinematic" and "potential" terms, respectively,

$$\rho_k \equiv \frac{1}{2}(\rho + p), \; \rho_v \equiv \frac{1}{2}(\rho - p). \qquad (35)$$

We will explore, for simplicity, the early universe only ($t \to 0$). Then, the evolution equations (2) can be cast under the form of a plane autonomous system, namely

$$\begin{cases} \dfrac{dx}{dN} = 3x \left\{ x - y + 1 - \left[ 1 + b(B_0 - 1) - 3dk^2 \right] \left( 1 + \tilde{\gamma} \dfrac{C\dfrac{\delta}{3} - \tilde{\theta}}{\tilde{\theta} + 1} \right) \right\} \\ \dfrac{dy}{dN} = 3 \left[ y(x - y + 1) - \left[ 1 - b(B_0 - 1) + 3dk^2 \right] \left( x + \tilde{\gamma} \dfrac{C\dfrac{\delta}{3} - \tilde{\theta}}{\tilde{\theta} + 1} y \right) \right]' \end{cases} \qquad (36)$$

where $\tilde{\theta} = b(B_0 - 1) - d(3\tilde{\gamma})^2$ and $N \equiv \ln a$.

To finish, setting $\dfrac{dx}{dN} = 0$ and $\dfrac{dy}{dN} = 0$ in Eqs. (34), we obtain the critical points in explicit form:

$$(x'_c; y'_c) = (0;0), \qquad (x''_c, y''_c) = \left( 0, 1 - \tilde{\gamma} \left[ 1 - b(B_0 - 1) + 3dk^2 \right] \dfrac{C\dfrac{\delta}{3} - \tilde{\theta}}{\tilde{\theta} + 1} \right). \qquad (37)$$

## 6. Conclusion

In the present paper we have investigated a coupled phantom/fluid cosmological model of dark energy which takes into account a fluid viscosity component and also dark matter in the FRW flat space-time. We have studied, in a number of different cases, the influence of the interaction between the dark energy and the dark matter contributions on the appearance of finite future time singularities of the different types listed in the literature. We have proven here, with the help of some simple examples, that the presence of dark matter can actually lead to a change of the singular behavior of the Hubble parameter (see also Ref. [33]). We have also obtained, in an explicit way, the critical points corresponding to the solution of the background equations.

Some remark is in order. It is very interesting to realize that inhomogeneous fluid cosmology can be often understood as modified gravity (for a review, see Ref. [36]), owing to the fact that it may be presented as a gravitational fluid with an inhomogeneous equation of state (see Ref. [37]). Then, it is plain that following the unification of inflation with dark energy in modified gravity (Refs. [36,38]) one can also achieve this unification in viscous cosmology, including the presence of dark matter. This will be further studied elsewhere.

**Acknowledgements**

This work was supported by a grant from the Russian Ministry of Education and Science, project TSTU-139 (A.V.T. and V.V.O.). E.E. has been partially supported by the project FIS2010-15640 (MINECO, Spain), by the CPAN Consolider Ingenio Project, and by AGAUR (Generalitat de Catalunya), contract 2009SGR-994.